\documentclass[3p]{elsarticle} \journal{Nuclear Physics B} 

\usepackage{amsmath}
\usepackage{mathtools}
\usepackage{amssymb}
\usepackage{graphicx}

\usepackage{tikz}
\usetikzlibrary{positioning}

\begin{document}
\begin{frontmatter}

\title{Shannon entropy and particle decays}


\author{Pedro Carrasco Mill\'an,  M. \'Angeles Garc\'{\i}a-Ferrero$^\dagger$,\\ Felipe J. Llanes-Estrada,  Ana Porras Riojano and Esteban M. S\'anchez Garc\'{\i}a}
\address{Departamento de F\'isica Te\'orica, Plaza de las Ciencias 2, Fac. CC. F\'isicas, Universidad Complutense de Madrid, 28040; \\
$^\dagger$ Instituto de Ciencias Matem\'aticas, C/ Nicol\'as Cabrera, 13-15, 28049;\\
Madrid, Spain}

\date{\today}

\begin{abstract}
We deploy Shannon's information entropy to the distribution of branching fractions 
in a particle decay. This serves to quantify how important a given new reported decay
channel is, from the point of view of the information that it adds to the already known
ones.
Because the entropy is additive, one can subdivide the set of channels and discuss, for example, how much information the discovery of a new decay branching would add; or subdivide 
the decay distribution down to the level of individual quantum states (which can be quickly counted by the phase space).
We illustrate the concept with some examples of experimentally known particle decay distributions.
\end{abstract}


\end{frontmatter}

\section{Introduction}

Shannon entropy~\cite{Shannon:1948zz} has found applications in all data-intensive fields of science; a recent review~\cite{Ma:2018wtw} with focus on heavy ion collisions provides an ample reference list and we refer the interested reader there.
This information entropy measures the uncertainty associated with a random variable, or 
when ignoring the value taken by the variable, of the average missing information content.

The decay width of an unstable particle can be decomposed into a sum over the partial widths
for each of its possible decay channels, $\Gamma=\sum\Gamma_j$. We can also characterize the decay by the branching ratios $BR_j=\Gamma_j/\Gamma$, and as
their sum is unity $\sum_j BR_j:=\sum_j \Gamma_j/\Gamma=1$,  they provide a probability distribution for the various decay channels ($j=1,\dots, N$).
 
This makes the information entropy of particle decay distributions a well posed observable to compute~\footnote{The quantity of Eq.~(\ref{Sdef}) is named entropy in analogy to the Gibbs mixing entropy, the increase in thermodynamic entropy $S=k_B \ln \Omega$ obtained upon mixing two gases totalling $N$ molecules, with partial molar fractions $x_1$ and $x_2$, 
\begin{equation*}
S=-k_B N\ \{x_1 \ln (x_1)+x_2\ln (x_2)\}
\end{equation*}
}
\begin{equation} \label{Sdef}
S=-\sum_{j
}BR_{j}\log BR_{j} \geq 0
\end{equation}
 and we will evaluate it with actual data from meson and gauge boson decays, all taken from~\cite{Patrignani:2016xqp}.

The maximum value of the entropy (for a decay distribution with a fixed number of channels) is
reached when they are all equally likely, that is,
$BR_j=\frac{1}{N}$ (because $\sum^{N}_{j=1}BR_j=1$ and $BR_i=BR_j$ for all $i,j$ in this case).
Then,
\begin{eqnarray} \label{maxEntropy}
S&\leqslant&-\sum_{i=j}^{N}BR_j\log(BR_j)\nonumber \\ 
&=&-\sum_{i=1}^{N}\frac{1}{N}\log(\frac{1}{N})=\log(N)
\end{eqnarray}

The minimum value is simply 0 and is reached when one channel concentrates all the probability,
$BR_1\simeq 1$, $BR_{j}\simeq 0, \, j>1$.
 Thus, we are looking at a variable $S\in(0,\log N)$ that characterises how disordered the decay products are, or namely, how difficult it is to predict the particular outcome of one decay event.

One can define the ``information'' function as the negative of the logarithm of the branching ratio, that is, $I_j :=-\log BR_j$. The entropy is then the average of that function over the distribution of decay channels,
\begin{equation}
\langle I_j \rangle = -\sum_j BR_j \log BR_j = S\ .
\end{equation}
Interpreting $I_j$ as the information obtained when a given particle decay proceeds through channel $j$ (a quantity associated to a given decay), $S$ is then the average information in the distribution of the random decay process (a quantity associated to all the decays, that is, to the decaying particle itself).

Shannon entropy has been used before in other contexts in particle physics. Early ones concentrated in the information entropy produced upon parton splittings ({\it e.g.} in jet emission)~\cite{Brogueira:1996fk,Cao:1996vc}. The concept has also been applied to study various fragment ratios after a heavy ion collision~\cite{Ma:2014vsa}.

Early-on after the Higgs boson discovery, d'Enterria~\cite{dEnterria:2012eip} observed that the Higgs sits in the window of maximum entropy of its decay distribution: were it heavier, around 200 GeV, above the $WW$ and $ZZ$ threshold, these two vector-boson channels would dominate the decay, with the rest of the Standard Model having much smaller branching fractions (and thus, the entropy being way smaller). As it is, at 125 GeV these boson channels are kinematically closed, and this makes the width small but the entropy large (as all the allowed Standard Model decays have to share a small portion of the decay).  

Alves, Dias and da Silva~\cite{Alves:2014ksa} then introduced a ``Maximum entropy principle'' elevating that observation about the Higgs to a more general principle, to try to predict the hypothetical axion mass~\cite{Alves:2017ljt,Alves:2017ipn}.

Even if the principle does not hold as a law of nature, the observation about the Higgs is sound and intriguing, and helps understand why its discovery happened so late in the development of high energy physics. 

In this article and in two companion proceedings publications~\cite{Llanes-Estrada:2017clj, Llanes-Estrada:2017ii}
we present numerous examples of computing the Shannon entropy of decaying mesons of multiple quark-flavor compositions, and of decaying electroweak bosons, and explore several new features of this variable and two related ones.

\section{Unknown decay channels }

\begin{figure}[h]
\begin{tikzpicture} 
  \node (img)  {\centerline{\includegraphics[width=0.45\columnwidth]{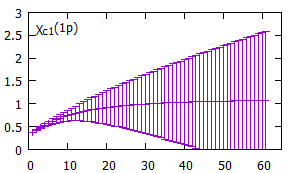}}};
  \node[below=of img, node distance=0cm, yshift=1cm,font=\large] {$N$};
  \node[left=of img, node distance=0cm, xshift= 4cm, rotate=90, anchor=center,yshift=-0.8cm,font=\large] {$S$};
 \end{tikzpicture}
		\caption{Shannon entropy for the decay distribution of the $\chi_{c1}(1p)$ charmonium as function of the number of channels included. Channels are added from left to right in order of decreasing branching fractions.
(A further plot that assesses the effect of removing radiative decays from the distribution can be found in~\cite{Llanes-Estrada:2017clj}). \label{fig:chic1}}
\end{figure}

\begin{figure}[h]
\begin{tikzpicture} 
  \node (img)  {\centerline{\includegraphics[width=0.45\columnwidth]{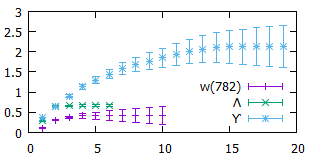}}};
  \node[below=of img, node distance=0cm, yshift=1cm,font=\large] {$N$};
  \node[left=of img, node distance=0cm,  xshift= 4cm, rotate=90, anchor=center,yshift=-0.8cm,font=\large] {$S$};
 \end{tikzpicture}

\caption{Shannon entropy for the decay distribution of mesons $\omega (782)$ and $\Upsilon (3s)$ (that can decay through the strong force), and for the $\Lambda$ baryon (that decays weakly). The $OX$ axis shows the number of channels included; the $OY$ axis, the resulting entropy. (A further plot showing the effect of removing the $\Upsilon(3s)$ radiative decay channels can be found in~\cite{Llanes-Estrada:2017clj}). \label{fig:3particulas}}
\end{figure}

In practice, many particles have complicated and multibody decays, so one does not always know the entire decay distribution. In that case,  
$\sum^N_{j=1} BR_j<1$ (with $BR_j< 1$). The discovery of new channels brings the sum closer to one, and the entropy increases. Nevertheless, if additional channels have a small branching fraction, their contribution to the entropy turns out to be negligible, and the entropy saturates.
This can be seen in figures~\ref{fig:chic1} and~\ref{fig:3particulas}. The error bars are computed from the experimental $\Delta BR_j$ uncertainties, and have been added linearly and not in quadrature, as determining different branching fractions of the same particle are often two very correlated measurements.

The first obvious approximation that we can perform is to bunch all unknown decay channels in just one with branching ratio equal to the missing part to reach 1 from the already known branching ratios at hand. The entropy is then, as a particular case of Eq.~(\ref{Sdef})
\begin{align}
S=&-\sum_{j_{\rm known}}BR_j \log BR_j \nonumber \\
&-\bigg(1-\sum_{j_{\rm known}}BR_j\bigg)\log\big(1-\sum_{j_{\rm known}}BR_j\big)\ .
\end{align}

In view of Eq.~(\ref{additivity}) below, this formula must underestimate the true entropy. 
As a way to estimate the error incurred, one could use perhaps the Kullback-Leibler divergence.
This states that the difference between the entropy and its estimate is given by
\begin{eqnarray}
S(BR_1,\dots, BR_M) -S(BR_1,\dots, BR_{N}, 1-\sum_{j=1}^{N} BR_j) = \nonumber \\
-\sum_{k=N+1}^M BR_k \log \bigg(\frac{BR_k}{\sum_{j=1}^{N} BR_j}\bigg) \nonumber \\ 
\end{eqnarray}
where $N$ of the $M$ channels are known and the number $M$ can be obtained from other considerations (for example, by studying which channels are open given phase space and conservation laws).
The $BR_k$, $N+1\leq k\leq M$ are the unknown branching ratios and $\sum_{j=N+1}^M BR_j=1-\sum_{j=1}^{N} BR_j$. For the purpose of the uncertainty estimate, they can be taken equal to each other, say
$BR_k=\frac{1-\sum_{j=1}^{N} BR_j}{M-N}$.

We will not pursue these theoretical error estimates any further, but content ourselves with propagating the experimental uncertainty $BR_j\pm \Delta BR_j$ to the entropy. 

\begin{figure}[h]
\begin{tikzpicture} 
  \node (img)  {\centerline{\includegraphics[width=0.6\columnwidth]{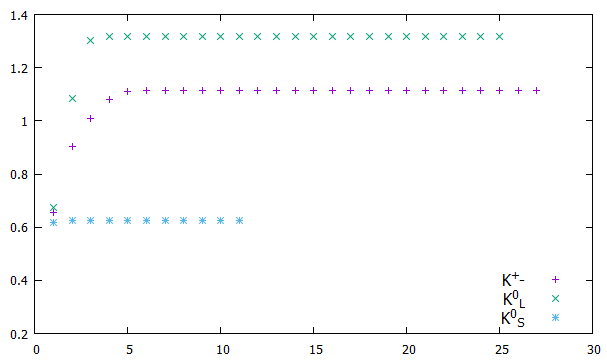}}};
  \node[below=of img, node distance=0cm, yshift=1cm] {$N$};
  \node[left=of img, node distance=0cm,  xshift= 3cm, rotate=90, anchor=center,yshift=-0.8cm] {$S$};
 \end{tikzpicture}
\begin{tikzpicture} 
  \node (img)  {\centerline{\includegraphics[width=0.6\columnwidth]{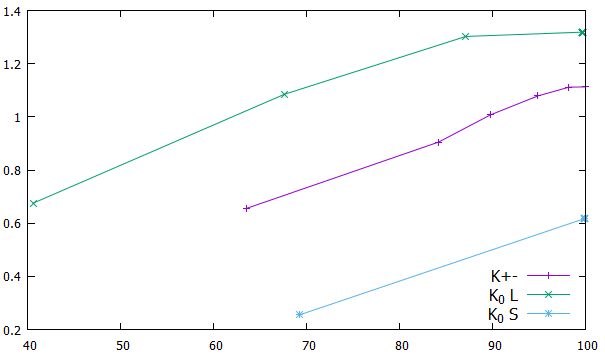}}};
  \node[below=of img, node distance=0cm, yshift=1cm] {$\sum_{j=1}^N BR_j  \,(\%)$};
  \node[left=of img, node distance=0cm,  xshift= 3cm, rotate=90, anchor=center,yshift=-0.8cm] {$S$};
 \end{tikzpicture}
\caption{\label{fig:kaones}
Shannon entropy for the decay product distribution of kaons. 
Top: $S$ as function of the number of channels included (in order of decreasing branching
fraction), analogous to figures~\ref{fig:chic1} and~\ref{fig:3particulas}. Bottom: $S$ as function of the percentage of the width accounted for the channels included (the seen channels correspond to the few points to the left of the top plot). Error bars are now not included for visibility. Further plots concentrating on $K^*_2(1430)$ have been presented in~\cite{Llanes-Estrada:2017clj}. 
}
\end{figure}
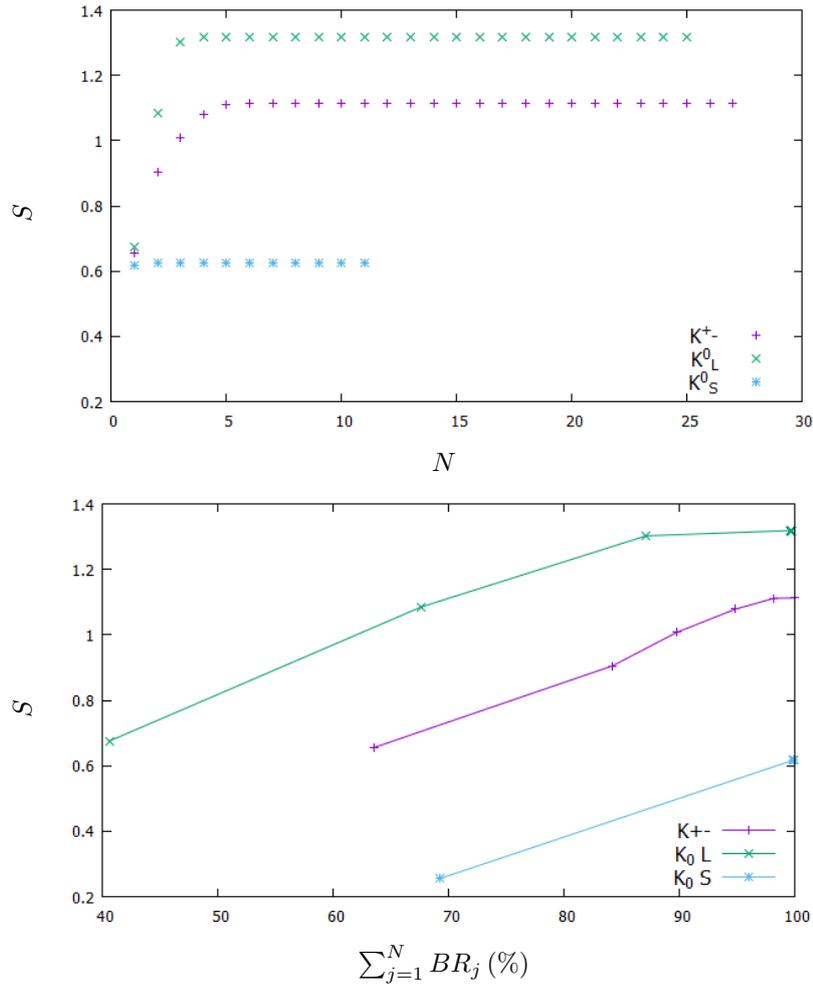

As seen in figures~\ref{fig:chic1}, ~\ref{fig:3particulas} and~\ref{fig:kaones}, the entropy increases monotonously upon adding more channels, but saturates into what seems at most logarithmic growth (which matches expectations from $\max(S)=\log N$). The second plot of figure~\ref{fig:kaones}  displays the entropy of certain kaon resonance decays not in terms of the number of channels included, but in terms of the sum of the branching ratios accounted up to the given channel, so the axis of abscissae ends at precisely 1.

\section{Entropy additivity and phase space}

There is a difficulty in analysing a given hadron decay chain: at which stage to count the final products. Does one have to descend all the way to stable particles, $p$, $e$ and $\nu$?  Is it sufficient to stay at the level of particles unable to decay strongly, so that $\pi$, $K$, etc. are considered final products? Or should one stop right away at the level of unstable hadron resonances with short lifetimes characteristic of the strong force of order $10^{-23}$ seconds? 

\begin{figure}[h]
\begin{tikzpicture} 
  \node (img)  {\centerline{\includegraphics[width=0.6\columnwidth]{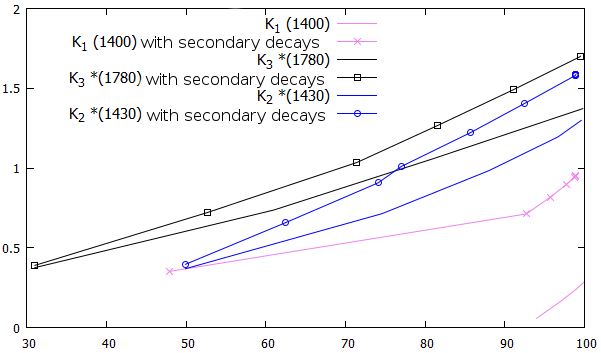}}};
  \node[below=of img, node distance=0cm, yshift=1cm] {$\sum_{j=1}^N BR_j  \,(\%)$};
  \node[left=of img, node distance=0cm,  xshift= 3cm, rotate=90, anchor=center,yshift=-0.8cm] {$S$};
 \end{tikzpicture}

\caption{Shannon entropy for the decay distribution of several kaon excitations, counting only primary decays to particles that are themselves unstable, or following the decay chain one more step to include secondary decays. \label{fig:secondarydecays}}
\end{figure}

Figure~\ref{fig:secondarydecays} shows an exercise where we study the decay of several excited kaons through another intermediate kaon and down to final products that are stable under the strong force, considering as an example the various subchannels to which the $K^*$ resonance decays to, {\it i.e.}
\begin{eqnarray*}
K_1(1400)\rightarrow K^*(892)\pi \rightarrow \left\{
\begin{array}{l}
 (K\pi )^\pm \pi \\ 
 (K\pi )^0 \pi \\ 
 (K^0\gamma )\pi\\
 (K^\pm \gamma) \pi \ .
\end{array}
\right.
\end{eqnarray*}

The entropy of the secondary decay chain (lines shown with hollow symbols) is quite different from that of the primary decay chain alone (only lines), since the secondary particles can decay through several channels, so a decision needs to be taken so as to how much to descend in the decay tree. For most applications in the strong interactions, when clear resonances can be identified one should stay at the first level (e.g. discount $\rho\pi$ from a given $\pi\pi\pi$ branching fraction).

Nevertheless, it is worth recalling a basic property of Shannon entropy~\footnote{This is a property that can be used to uniquely determine the entropy function, together with continuity and with monotonicity for the particular $BR_j \propto 1/j$ branching fraction distribution, which leads to a $\sum_j BR_j \log BR_j$ formula.},
namely its additivity. 

If each of the branching ratios at the first level, $BR_j$, counts the joint probability of routing the decay into certain subchannels $j_1,j_2,\dots, j_{m_j}$, the entropy at the second level, which includes all  subchannels, can be obtained from the entropy at the first level in terms of the $BR_j$ and the entropy of each subdivision as follows~\footnote{The second term is familiar from the usual thermodynamic entropy: if any molecule can be in two gas volumes 1 and 2 with probabilities $p$ and $1-p$, the gas entropy satisfies $S(1\cup 2)= p S(1) + (1-p) S(2)+S(p, 1-p)$, where
$ S(p,1-p) = -p\log p - (1-p) \log(1-p)$.},
\begin{equation} \label{additivity}
S = S(BR_1,BR_2,\dots, BR_N) + \sum_{j=1}^N BR_j S\left(\frac{BR_{j_1}}{BR_j}, \dots, \frac{BR_{j_{m_j}}}{BR_j}  \right) \ .
\end{equation}

The first function is the entropy at the higher level (where the subchannels are all bunched in one) and the second is a sum, over each of the primary channels, of the entropy within each of them weighted with its overall probability $BR_j$. That second term explains the difference between the lines with and without symbols in figure~\ref{fig:secondarydecays}.

\begin{figure}[h]

\begin{tikzpicture} 
  \node (img)  {\centerline{\includegraphics[width=0.6\columnwidth]{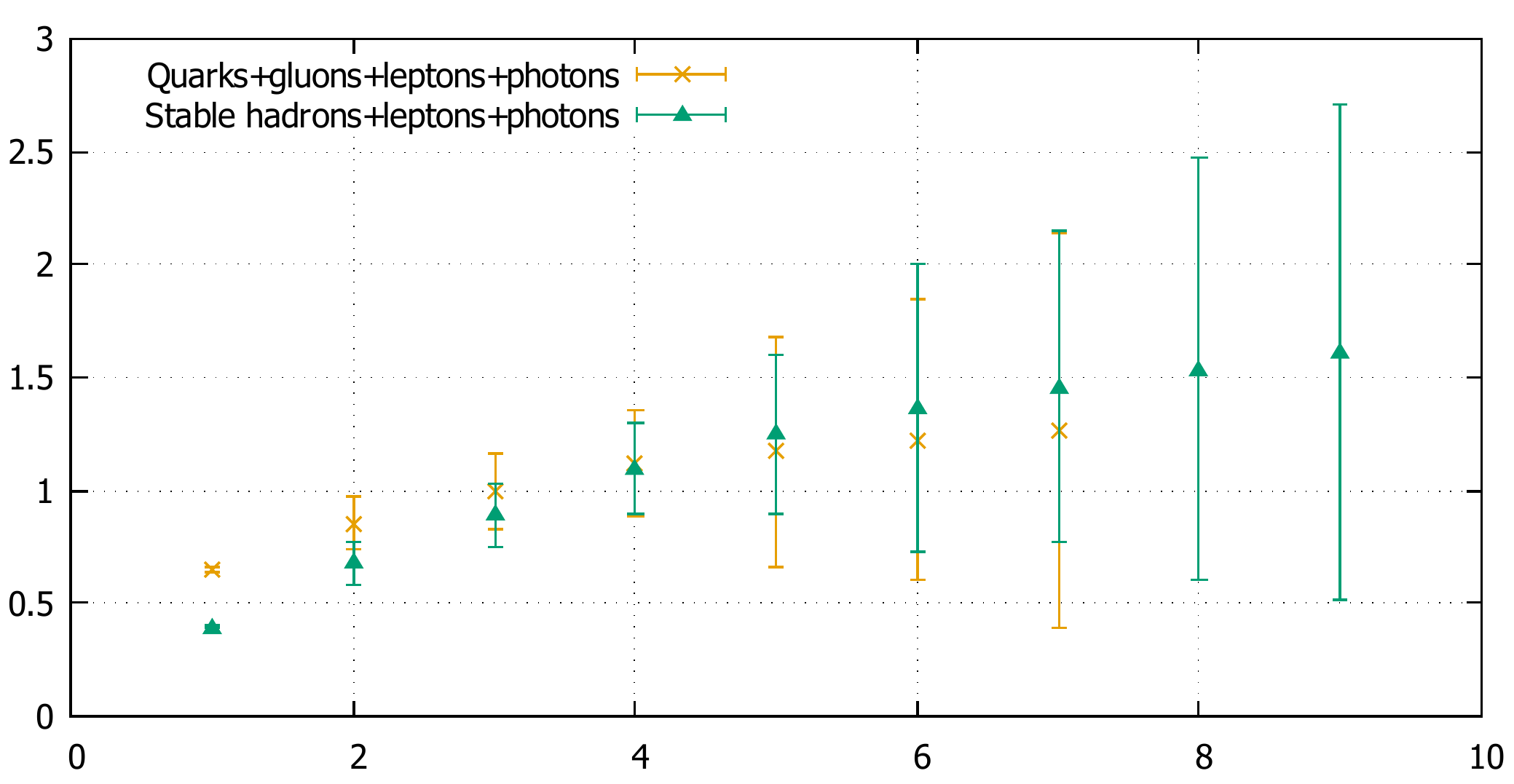}}};
  \node[below=of img, node distance=0cm, yshift=1.1cm,font=\large] {$N$};
  \node[left=of img, node distance=0cm,  xshift= 3cm, rotate=90, anchor=center,yshift=-0.8cm,font=\large] {$S$};
 \end{tikzpicture}

\caption{Shannon entropy for the decay distribution of the $J/\psi$ charmonium into final states (where the hadron species are enumerated) and into intermediate states (where the description is kept at the level of quarks and gluons). Within the experimental errors, it is not easy to ascertain the difference between the two entropies. \label{fig:Jpsitwochains}}
\end{figure}

Quite surprisingly, the decay entropies are not always so different. This is highlighted by the 
decay entropy of the $J/\psi$ in figure~\ref{fig:Jpsitwochains}.
In that case, the entropy accumulated when describing the decay in terms of intermediate quarks and gluons is similar to that accumulated when employing the identified final state hadrons. The later one is a bit larger, consistently with Eq.~(\ref{additivity}), but the experimental errors propagated from the uncertainty in the branching fractions are much larger than the difference. 

As a matter of principle, the lowest level to which one can descend in a decay chain is that of individual quantum states, to which we turn next.

\subsection{Entropy in terms of phase space}
The partial decay width can be written in terms of the invariant Feynman amplitude
\begin{equation}
	d\Gamma_{i\rightarrow f}=\frac{1}{2M_i}\overline{\arrowvert \mathcal{M}\arrowvert^2}d\rho
\end{equation}
and the Lorentz Invariant Phase Space
\begin{eqnarray}
	d\rho =\ \ \ \ \ \ \ \ \ \ \ \ 
 \\ \nonumber (2\pi)^4 \delta^{(3)} \left(\sum_{f=1}^n\overrightarrow{p_f}-\overrightarrow{P_i}\right)\delta \left(\sum_{f=1}^n E_f- E_i\right) \prod_{f=1}^{n}\frac{dp_f}{(2\pi)^32E_f}
\end{eqnarray}
which counts the number of available quantum states (thus, the decay distribution cannot be subdivided any further). Restricting ourselves to two--body decay channels, the integrated phase space is
\begin{equation} 
\rho (E) = \frac{(2\pi)^4}{(2\pi)^6} \int \int \frac{d^3 p_1}{2E_1} \frac{d^3 p_2}{2E_2}\delta ({\overrightarrow p}_1+{\overrightarrow p}_2-{\overrightarrow P}) \delta (E_1 + E_2 -E) 
\end{equation}
that, with center of mass kinematics, yields the well-known relation
\begin{equation}
\rho (E) = \frac{1}{4\pi}\frac{[(E^2-(m_2-m_1)^2)(E^2-(m_2+m_1)^2)]^{\frac{1}{2}}}{2E^2}\ .
\end{equation}

\begin{figure}[h]
\begin{center}
\begin{tikzpicture} 
  \node (img)  {\centerline{\includegraphics[width=0.6\columnwidth]{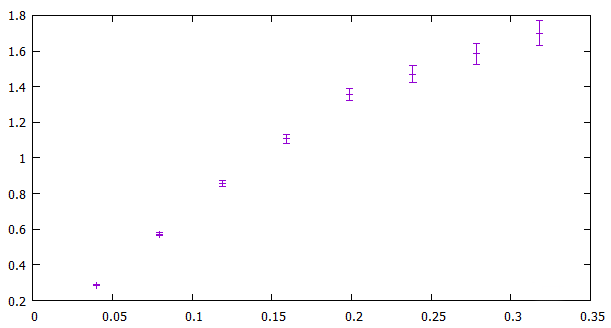}}};
  \node[below=of img, node distance=0cm, yshift=1cm] {$\sum_j\rho_j$};
  \node[left=of img, node distance=0cm,  xshift= 3cm, rotate=90, anchor=center,yshift=-0.8cm,font=\large] {$S$};
 \end{tikzpicture}
\caption{Entropy as function of the accumulated phase space $\sum_{j=1}^N \rho_j$ upon including successive decay two--body channels of the electroweak $Z$ boson, ordered from larger to smaller branching fraction.
\label{fig:Z}}
\end{center}
\end{figure}

We plot the entropy for the decay distribution of the electroweak $Z$ boson against the accumulated phase-space in figure~\ref{fig:Z} (the case of the $W$ boson has also been analysed and reported in \cite{Llanes-Estrada:2017clj}).

The $OY$ axis should now be considered as arbitrarily normalised, as we are still plotting the entropy $S(BR_1,\dots, BR_N)$ out terms correcting for the internal entropy of each channel. Nevertheless, the $OX$ axis is now scaled with the correct phase space for each of the included  two-body decays. There is not much qualitative difference at this point (a vertical stretching of the entropy function if we took into account the internal entropy), so we will continue plotting entropy at an aggregate level channel by channel.

\section{Information value of discovering a new decay channel}

We wish to propose  a simple criterion to quantify what information the discovery of a new branching fraction provides to the knowledge of a particle decay (purely from the statistical point of view, without entering to judge whether that decay may be showing the violation of an approximate symmetry, or be a golden mode for certain observables or any other qualitative effects that need to be judged on a case by case basis). 

\begin{figure}[h]
\begin{center}
\begin{tikzpicture} 
  \node (img)  {\centerline{\includegraphics[width=0.45\columnwidth]{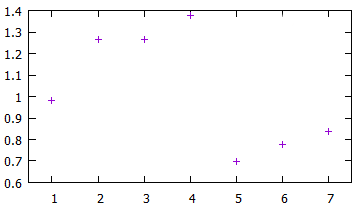}}};
  \node[below=of img, node distance=0cm, yshift=1.1cm,font=\large] {$N$};
  \node[left=of img, node distance=0cm,  xshift= 4cm, rotate=90, anchor=center,yshift=-0.7cm,font=\large] {$\frac{\Delta S(N)}{\Delta \log(N)}$};
 \end{tikzpicture}

\caption{Increase of the normalized entropy for the decay distribution of the $Z$ boson 
as function of the number of channels.\label{fig:normincrement}}
	\end{center}
\end{figure}

The first obvious effect is that of Eq.~(\ref{maxEntropy}).
 Having observed that the maximum entropy grows as the log of the number of channels, if all were equally weighted, the actual importance of a new channel can be obtained by studying the separation of the entropy from this maximum value. 
Therefore, we propose two possible measures of this added information. One is the normalized entropy increment, defined by
\begin{equation}
\frac{\Delta S(N)}{\Delta \log(N)}:=\frac{S(N+1)-S(N)}{\log(N+1)-\log(N)}
\end{equation}
that is plotted in figure~\ref{fig:normincrement}, where we show how this normalized entropy increment would evolve upon sequentially including (eventually, \emph{discovering}) the represented decay channels of the $Z$ boson. ($N$ increases by one every time a new channel is added to the list.)

In  the figure we see that the discovery of decay channels 5, 6 and 7 would then be less significant, from the point of view of information theory, than the discovery of one of the channels 1 through 4.

\begin{figure}[h]
\begin{center}
\begin{tikzpicture} 
  \node (img)  {\centerline{\includegraphics[width=0.45\columnwidth]{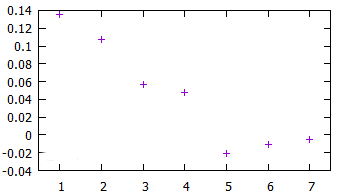}}};
  \node[below=of img, node distance=0cm, yshift=1.1cm,font=\large] {$N$};
  \node[left=of img, node distance=0cm,  xshift= 4cm, rotate=90, anchor=center,yshift=-0.7cm,font=\large] {$\Theta$};
 \end{tikzpicture}

\caption{Change in the degree of likeness $\Theta$ upon discovering each new channel (so that the last unknown channel splits off part of its probability to that new known one, and $N$ increases by one unit) for the $Z$ boson.\label{fig:likeness}}
	\end{center}
\end{figure}

Another possibility is to employ a certain ``degree of likeness'' (to the maximum possible entropy) which can be simply defined by
$\frac{S(N)}{\log(N)}\in (0,1)$. Its increment upon adding one new channel would then be
\begin{equation}
\Theta := \frac{S(N+1)}{\log(N+1)}-\frac{S(N)}{\log(N)}\ .
\end{equation}
A positive $\Theta$ means that the entropy of distribution steps closer to the maximum possible value of $S$ upon introducing the new channel; this can happen when the new channel has a branching fraction similar to the ones already known. If $\Theta$ is negative, the entropy decreases relative to its maximum possible value, and
the new channel is very dissimilar from the others (typically smaller). This function, applied to the same decay distribution of the $Z$ boson as in figure~\ref{fig:normincrement} is plotted in figure~\ref{fig:likeness}.

We can ascertain once more in this figure that channels 5, 6 and 7 contribute less to the entropy because their branching fractions are much smaller than those of the channels included earlier.

\section{Base of the logarithm}

The base of the logarithm in Eq.~(\ref{Sdef}) provides a mean to compare different decay chains. Very often in computer science, 
$\log_k$ is taken with $k=2$ so that information is measured in \emph{bits}. Our entropy here has been rather presented in \emph{nats} by employing the natural logarithm. 
An interesting additional choice is to use $k=N$, the actual number of channels
needed to describe the particle's decay. (For small $N$ it is worth noting that, since we are usually packaging an unspecified number of unknown channels into an additional one, 
we will rather use $k=N+1$).

This choice of scaling the base with $N$ has an advantage to compare the entropy of decay distributions for different particles associated, not to the number of channels, but rather to the inhomogeneity of the decay product distribution among them. This comes about because the maximum possible value of the entropy is then $\log_N N =1$ (or, being precise, $\log_{N+1} (N+1)=1$) and one can then obviously compare two particles on the same scale.

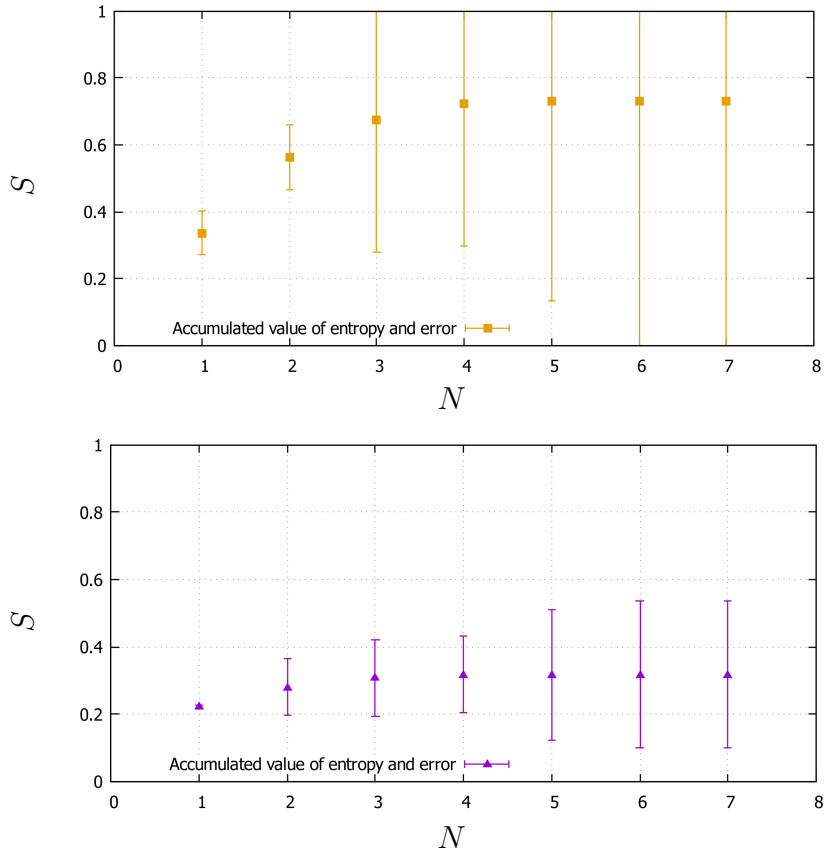
\begin{figure}[h]
\begin{tikzpicture} 
  \node (img)  {\centerline{\includegraphics[width=0.6\columnwidth]{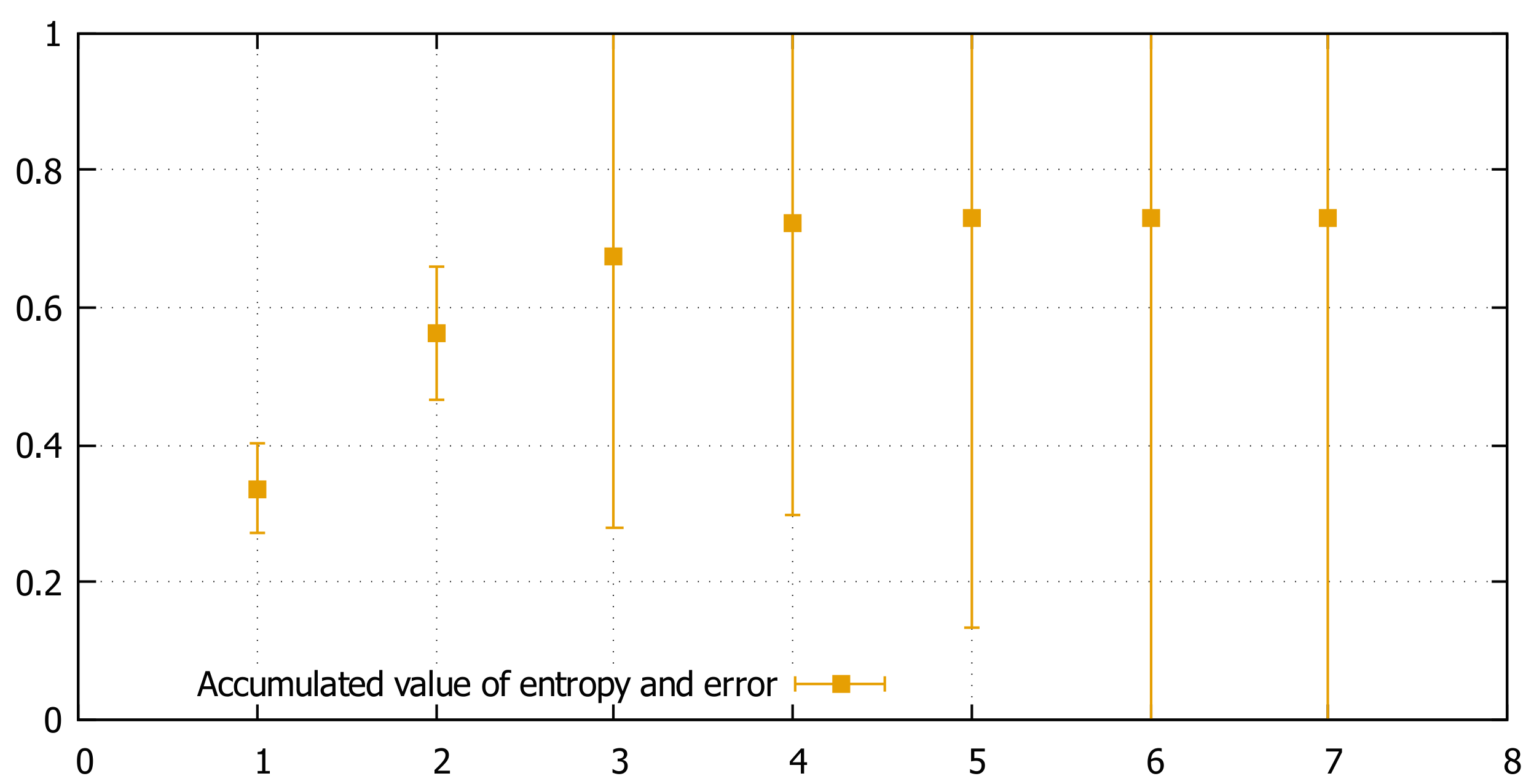}}};
  \node[below=of img, node distance=0cm, yshift=1.1cm,font=\large] {$N$};
  \node[left=of img, node distance=0cm,  xshift= 3cm,rotate=90, anchor=center,yshift=-0.7cm,font=\large] {$S$};
 \end{tikzpicture}
\begin{tikzpicture} 
  \node (img)  {\centerline{\includegraphics[width=0.6\columnwidth]{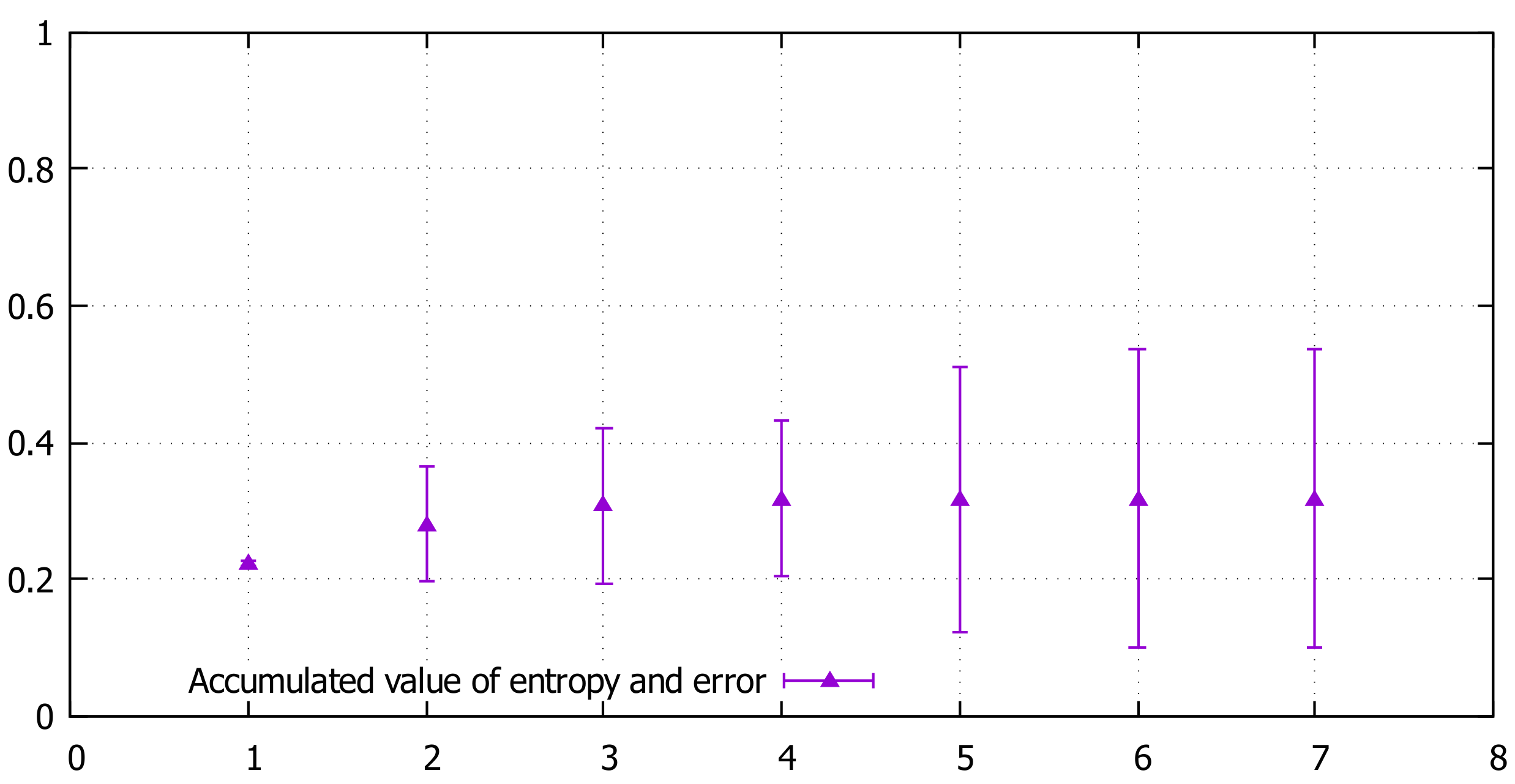}}};
  \node[below=of img, node distance=0cm, yshift=1.1cm,font=\large] {$N$};
  \node[left=of img, node distance=0cm,  xshift= 3cm, rotate=90, anchor=center,yshift=-0.7cm,font=\large] {$S$};
 \end{tikzpicture}

\caption{\label{fig:f1f2}
Shannon entropy for the decay product distribution of two example flavor-singlet $f$ mesons against the number of channels included in the decay (analogous to figure~\ref{fig:3particulas}). The base of the logarithm is here chosen to be $N+1$, the number of decay channels included. 
}
\end{figure}

The comparison is even more telling if both particles have the same number of relevant decay channels. This is approximately the case for instance for the pair of $f_1(1285)$, an axial $J^{PC}=1^{++}$ meson, and its multiplet partner $f_2(1270)$, a tensor $2^{++}$ meson. This last one has a decay very much dominated by $\pi\pi$ (85\%) while the former has the probability more distributed among the $\eta\pi\pi$, $4\pi$, $K\bar{K}\pi$ and $a_0(980)\pi$ channels, the rest being minor. Figure~\ref{fig:f1f2} shows the data; in both cases the entropy has its maximum possible value at 1.

\begin{figure} [h]
\begin{tikzpicture} 
  \node (img)  {\centerline{\includegraphics[width=0.6\columnwidth]{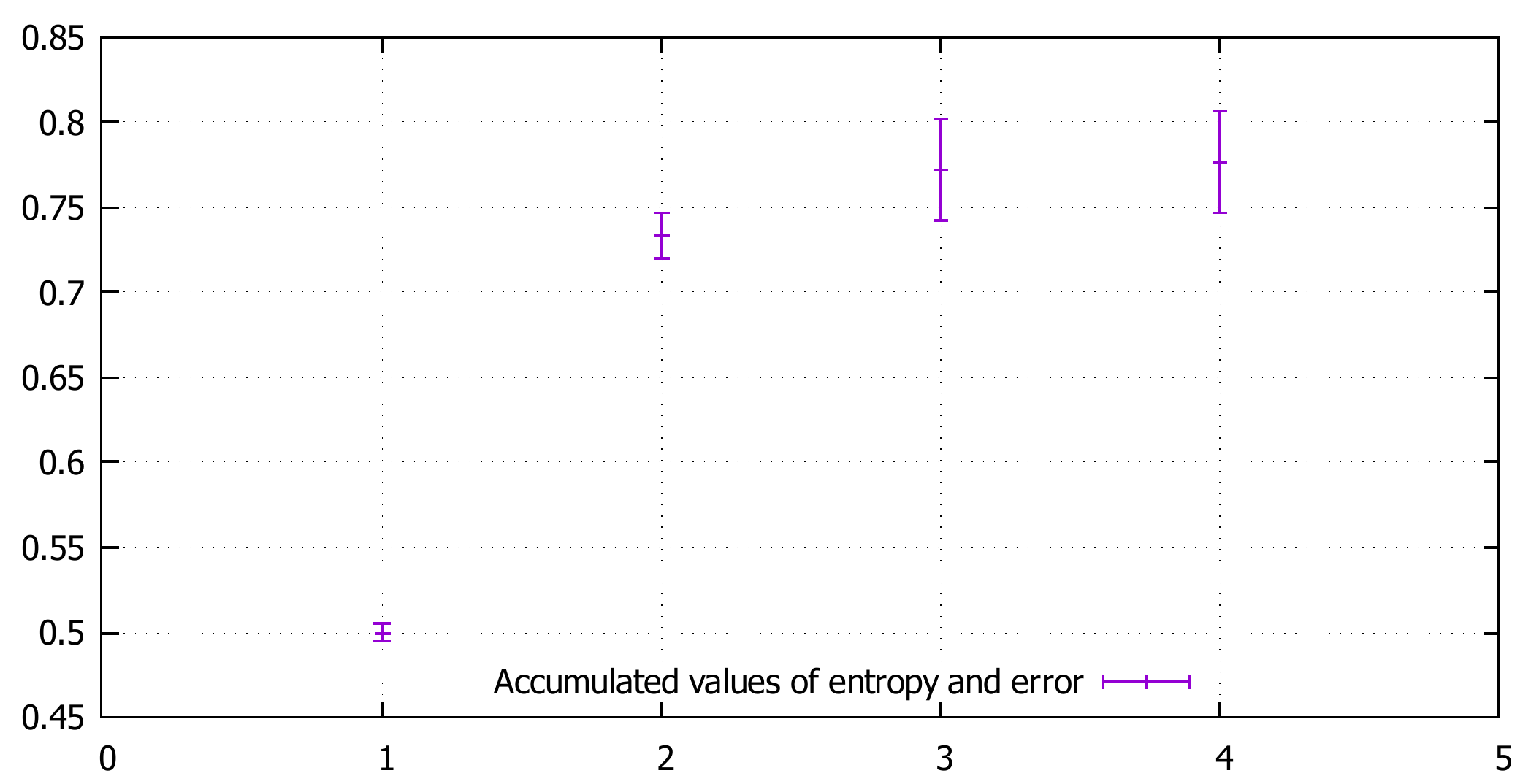}}};
  \node[below=of img, node distance=0cm, yshift=1.1cm,font=\large] {$N$};
  \node[left=of img, node distance=0cm,  xshift= 3cm, rotate=90, anchor=center,yshift=-0.7cm,font=\large] {$S$};
 \end{tikzpicture}
\begin{tikzpicture} 
  \node (img)  {\centerline{\includegraphics[width=0.6\columnwidth]{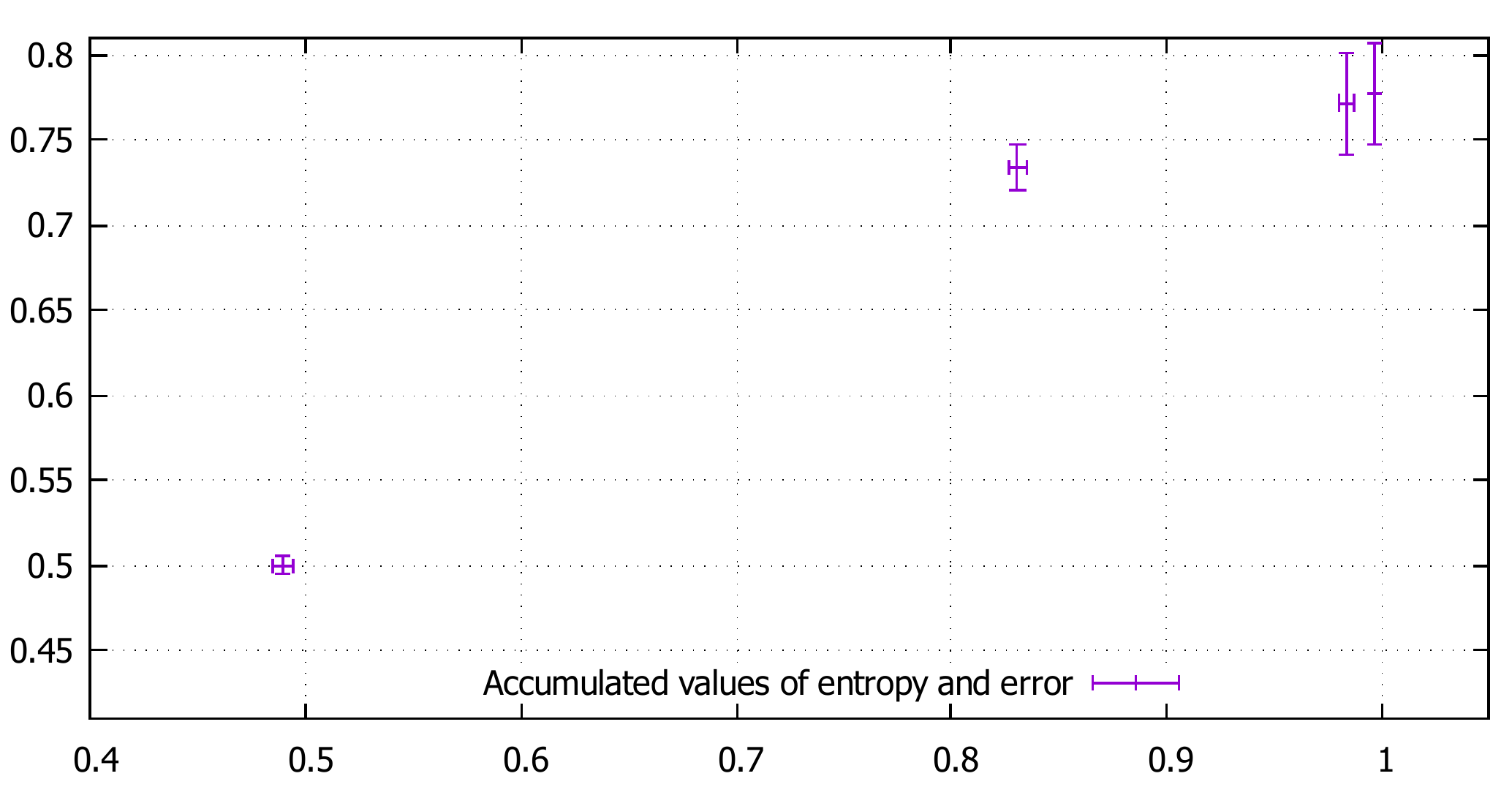}}};
  \node[below=of img, node distance=0cm, yshift=1.1cm] {$\sum_{j=1}^N BR_j$};
  \node[left=of img, node distance=0cm,  xshift= 3cm, rotate=90, anchor=center,yshift=-0.7cm,font=\large] {$S$};
 \end{tikzpicture}

\caption{\label{fig:phi}
Shannon entropy for the decay product distribution of the $\phi$ meson (analogous to figure~\ref{fig:kaones} but here the base of the logarithm is chosen to be $N+1$). 
Top: $S$ as function of the number of channels included (in order of decreasing branching
fraction). Bottom: $S$ as function of the percentage of the width accounted for.
}
\end{figure}
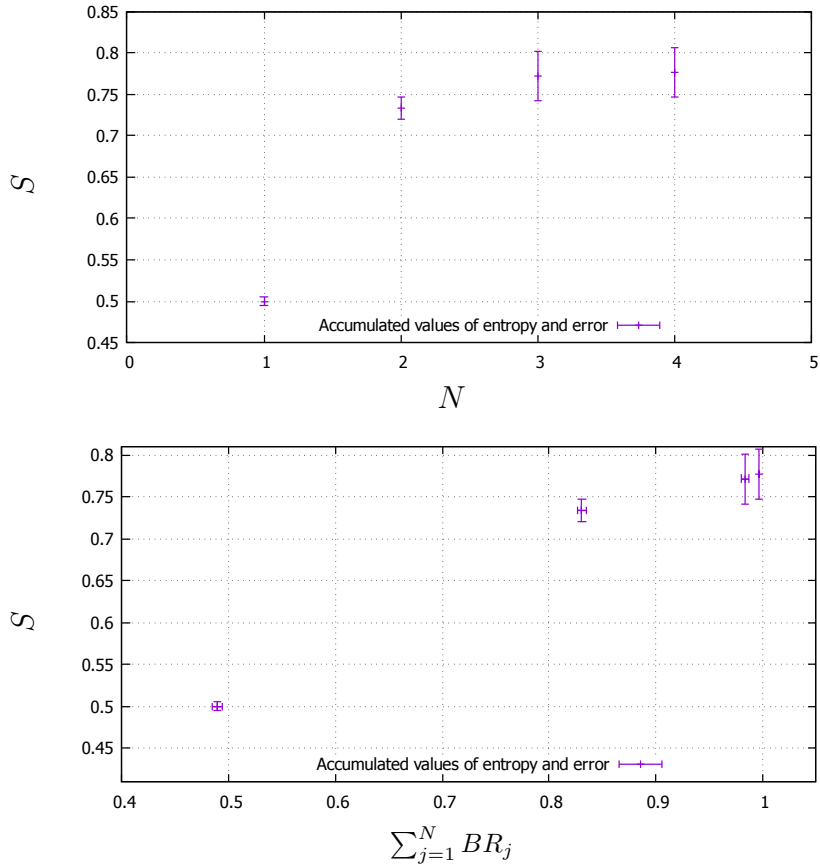

As a second example, figure~\ref{fig:phi} shows the entropy of the decay distribution of the $\phi$ meson, also normalised to 1 by choosing $\log_N BR_j$ instead of the natural logarithm. 

\section{Further observations and outlook}

\begin{figure}[h]
\begin{tikzpicture} 
  \node (img)  {\centerline{\includegraphics[width=0.45\columnwidth]{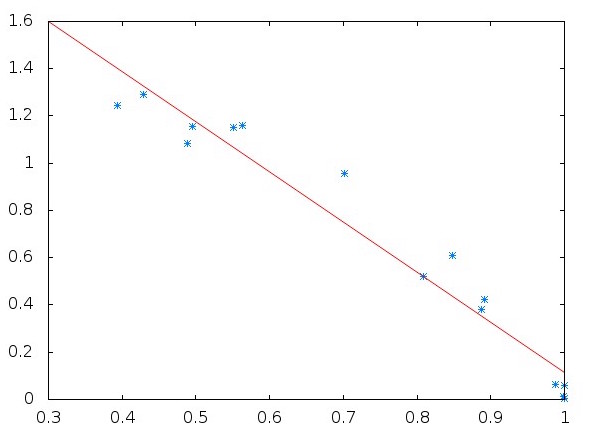}}};
  \node[below=of img, node distance=0cm, yshift=1.1cm] {$\max_j BR_j$};
  \node[left=of img, node distance=0cm,  xshift= 4cm, rotate=90, anchor=center,yshift=-0.7cm,font=\large] {$S$};
 \end{tikzpicture}

\caption{\label{fig:correlation}
We show the clear anticorrelation between the information entropy $S$ and the maximum value of the branching fractions (for the light, hidden-flavor mesons). The entropy is obviously very small if one certain channel dominates the decay.}
\end{figure}

Figure~\ref{fig:correlation} plots, for the light, unflavored mesons ($\eta$, $\omega$, $\phi$, etc.) the entropy against the maximum of the branching fractions $BR_j$ for the various decay channels of each. 

There is a clear anticorrelation between the two variables: the entropy (lack of predictivity about any one particular decay) is much larger when there is no dominant decay branching fraction, as should be evident to the reader. Thus, it is more informative to discover a new decay channel when none carries a fraction close to unity of the total decays.

A further observation is that, not uncommonly, the branching fractions are ordered in a geometric hierarchy $BR_0$, $BR_1=f\ BR_0,\dots, $ $BR_N=f^N BR_0$ with $f<1$ (this is typically seen in entropy functions that grow quickly for the very first channels and saturate almost immediately; a plot for such a distribution with $f=1/2$ has been relegated to the proceedings in~\cite{Llanes-Estrada:2017clj}). This statement is reflected in the following approximation~\footnote{In an extreme, idealized case, $N\to\infty$ and $BR_0\sum_{j=0}^\infty f^j = \frac{BR_0}{1-f}=1$, so that $f$ and $BR_0$ are not independent; but this is not the case in practical examples that only approximately follow this rule and where not all channels are known.},
\begin{eqnarray}
S = - BR_0 \sum_{j=0}^N f^j \left( j\log f + \log BR_0   \right) \ .
\end{eqnarray}
It converges quickly for many particles (typically those with few open strong-decay channels)
but $f$ does depend on the particle in question.

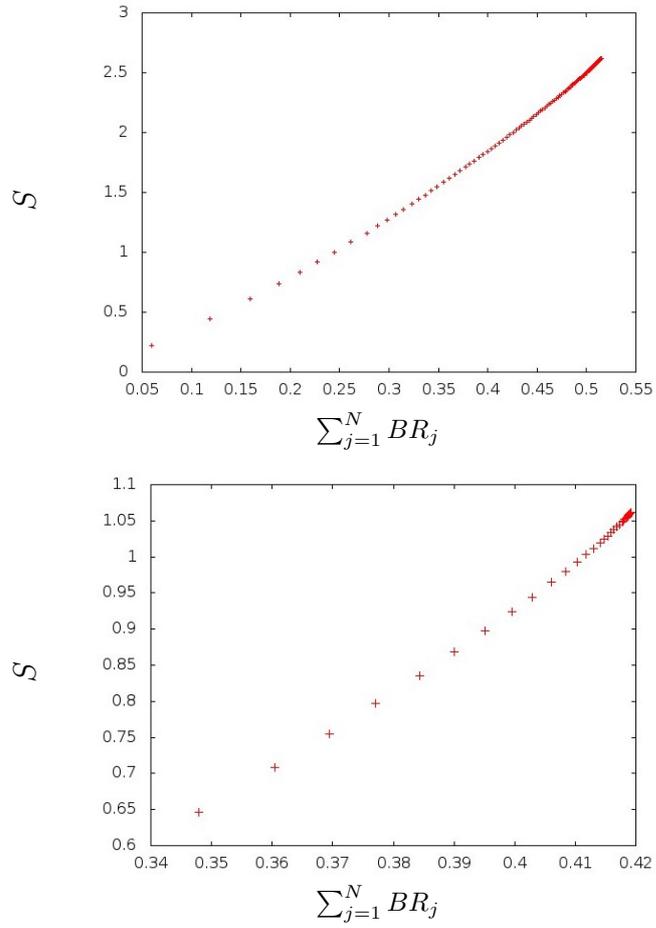
\begin{figure}[h]
\begin{tikzpicture} 
  \node (img)  {\centerline{\includegraphics[width=0.45\columnwidth]{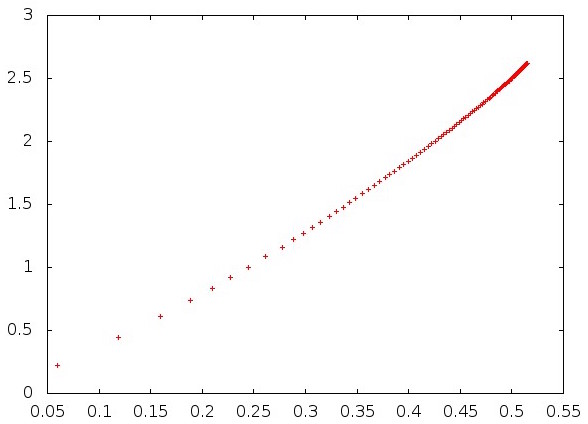}}};
  \node[below=of img, node distance=0cm, yshift=1.1cm] {$\sum_{j=1}^N BR_j$};
  \node[left=of img, node distance=0cm,  xshift= 4cm, rotate=90, anchor=center,yshift=-0.7cm,font=\large] {$S$};
 \end{tikzpicture}
\begin{tikzpicture} 
  \node (img)  {\centerline{\includegraphics[width=0.45\columnwidth]{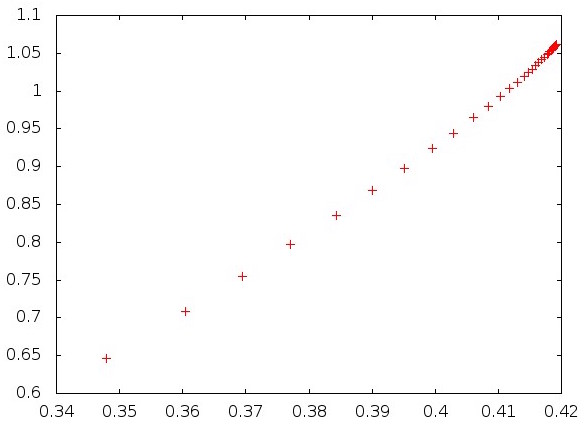}}};
  \node[below=of img, node distance=0cm, yshift=1.1cm] {$\sum_{j=1}^N BR_j$};
  \node[left=of img, node distance=0cm,  xshift= 4cm, rotate=90, anchor=center,yshift=-0.7cm,font=\large] {$S$};
 \end{tikzpicture}

\caption{\label{fig:charmonio} Entropy for the vector $J/\psi$ (top plot) and axial vector $\chi_{c1}$  (bottom plot) charmonia, against the accumulated branching ratio accounted for. }
\end{figure}

\begin{figure}
\begin{tikzpicture} 
  \node (img)  {\centerline{\includegraphics[width=0.45\columnwidth]{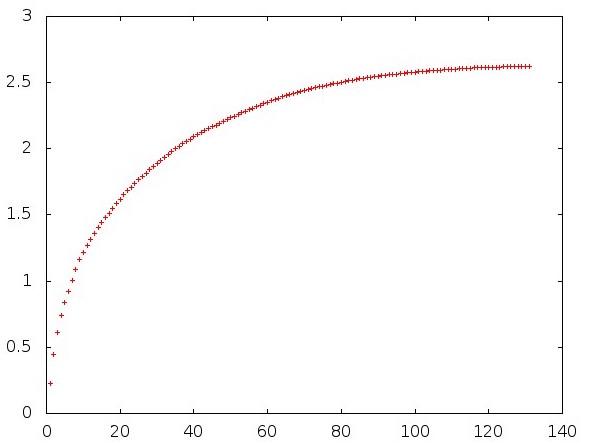}}};
  \node[below=of img, node distance=0cm, yshift=1.1cm] {$N$};
  \node[left=of img, node distance=0cm,  xshift= 4cm, rotate=90, anchor=center,yshift=-0.7cm,font=\large] {$S$};
 \end{tikzpicture}
\begin{tikzpicture} 
  \node (img)  {\centerline{\includegraphics[width=0.45\columnwidth]{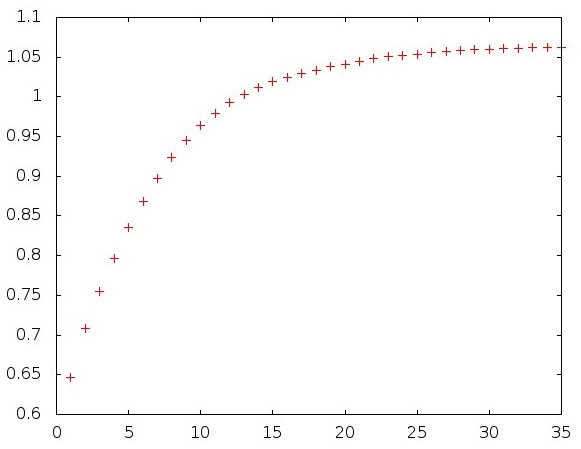}}};
  \node[below=of img, node distance=0cm, yshift=1.1cm] {$N$};
  \node[left=of img, node distance=0cm,  xshift= 4cm, rotate=90, anchor=center,yshift=-0.7cm,font=\large] {$S$};
 \end{tikzpicture}

\caption{\label{fig:charmonio2} Same as in figure~\ref{fig:charmonio} but plotted as a function of the number of channels instead of the accumulated branching fraction.  Because many channels have commensurable branching fractions, $S$ grows approximately linearly with the number of channels until $\sum_i BR_i$ starts being a sizeable fraction of 1. The characteristic $\log N$ behavior is visible following that linear regime.}
\end{figure}

Mesons that contain heavy quarks but low excitation number do not fall in this category. Instead, they possess many channels (with light valence quarks only) that have similar branching ratios. Then the entropy function grows linearly with the number of channels and many of them are required to start saturating it. This is best visible in figures~\ref{fig:charmonio} and~\ref{fig:charmonio2}, especially the second one (entropy against the number of channels).

For those two low-lying charmonia, about half of the total width is accounted for. Each new channel, of a size similar to those previously known, increases the entropy practically in proportion to its branching fraction (figure~\ref{fig:charmonio}). The characteristic $\log N$ growth of the entropy is however visible if we plot the same data against the number of channels instead of the branching fraction accounted for (figure~\ref{fig:charmonio2}).

To conclude, we have found that Shannon's entropy is an interesting tool to ascertain the relative importance of different decays. Taking into account the sheer size of the particle physics decay data collected by the community and ordered by the Particle Data Group, this and other methods of information theory find a rich field of applicability.

As we have seen in numerous examples, the generic behaviour of the entropy of the distribution against the number of channels is a linear increase for the first few, larger ones, followed by a saturation well below the entropy's maximum for $N$ channels, $\log N$.
 
We have discussed how to compare different particles, using the logarithm of base $N$ is fair as it normalises the maximum entropy to unity.
We have also discussed simple derived functions that help quantify the amount of entropy that a given decay channel adds to the distribution after its discovery.

And finally, we have shown the anticorrelation between the entropy and the maximum branching fraction of any decay channel. Shannon's entropy is maximized by particles that decay more or less equally through their decay channels (perhaps because the decaying particle is below the threshold of the channel it couples more strongly to).

\newpage
\section*{Acknowledgments}
Work supported by grants from MINECO FPA2014-53375-C2-1-P and FPA2016-75654-C2-1-P, ERC Starting Grant~633152, ICMAT-Severo Ochoa project SEV-2015-0554, and carried out in the inspiring atmosphere of the theoretical physics department and UPARCOS. M.A.G.F. acknowledges the Spanish MINECO for a Severo Ochoa FPI contract. 

\newpage

\end{document}